\documentclass[useAMS,usenatbib]{mn2e}
\usepackage{epsfig}
\usepackage{multirow}
\usepackage{aas_macros}     
\usepackage{amsfonts}
\title[Testing dark energy using pairs of galaxies in redshift space]
{ Testing dark energy using pairs of galaxies in redshift space
}
\author[E.~Jennings, C.~M.~Baugh, S.~Pascoli]
{E. Jennings$^{1,2}$\thanks{E-mail: elise.jennings@durham.ac.uk}, C. M. Baugh$^{1}$, S. Pascoli$^{2}$\\
$^{1}$ Institute for Computational Cosmology, Department of Physics, Durham University, South Road, Durham, DH1 3LE, U.K.\\
$^{2}$ Institute for Particle Physics Phenomenology, Department of Physics, Durham University, South Road, Durham, DH1 3LE, U.K.\\
}
\begin{document}

\date{}

\pagerange{\pageref{firstpage}--\pageref{lastpage}} \pubyear{2011}

\maketitle

\label{firstpage}

\begin{abstract}
The distribution of angles subtended between pairs of
galaxies and the line of sight,
which is uniform in real space, is distorted by 
their peculiar motions, and has been
proposed as a probe of cosmic expansion. We test this idea
using N-body simulations of structure formation in a cold
dark matter universe with a cosmological constant and in two
variant cosmologies with different dark energy models. We
find that the distortion of the distribution of angles is
sensitive to the nature of dark energy. However, for 
the first time, our simulations also reveal dependences of
the normalization of the distribution on both redshift and
cosmology that have been neglected in previous work. This
introduces systematics that severely limit the usefulness of
the original method. Guided by our simulations, we devise a new,
improved test of the nature of dark energy. We demonstrate that 
this test does not require prior knowledge of the background 
cosmology and that it can even distinguish between models that have 
the same baryonic acoustic oscillations and dark matter halo
mass functions. Our technique could be applied to the completed BOSS
galaxy redshift survey to constrain the expansion history
of the Universe to better than 2\%. The method will also produce different
signals for dark energy and modified gravity cosmologies even  when they
have identical expansion histories, through the different peculiar
velocities induced in these cases.
\end{abstract}


\section{Introduction}

One of the primary scientific goals of ongoing and future
galaxy surveys is to determine what is responsible for
the accelerating expansion of the Universe
\citep{EUCLID, 2007AAS...21113229S,2009arXiv0904.0468S,2010MNRAS.tmp..776B,2011arXiv1106.1706S}.
There are two main considerations which affect  current tests of dark energy. Firstly, the differences between the observables expected from competitive cosmological models
are small. Secondly, given the huge volumes that will be covered by future surveys it is likely that systematic errors will dominate the interpretation of the measurements.
For both these reasons it is generally accepted that the dark energy challenge should be
tackled using multiple cosmological probes \citep{2006ewg3.rept.....P, 2006astro.ph..9591A}.
Guided by numerical simulations, we assess a recently proposed test, a version of the Alcock-Paczynski test \citep{1979Natur.281..358A}, which
uses measurements of galaxy pairs to constrain the cosmological
model. We expand the available probes of dark energy by setting out an improved
version of the test which we show can distinguish models that otherwise cannot be
separated by existing methods.

The Alcock-Paczynski test measures the distortion of a spherical object assuming an incorrect cosmological model is used to compute distances. The version of 
this test considered here models the distortion in a spherically symmetric distribution of galaxy pair angles in redshift space and was first proposed in a form similar
to that used in this paper by \citet{1994MNRAS.269.1077P}, who considered the distribution of
angles between quasar pairs.
Recently \citet{2010Natur.468..539M} introduced an important revision to this test by
considering the angle between pairs of galaxies viewed in redshift space.
Building on the work of  Marinoni \& Buzzi, the method outlined in this paper is a refined geometrical test of dark energy. 
The critical feature of our extension is the use of N-body simulations of
different dark energy models to test the idea that measuring the anisotropic distribution of
galaxy pairs in redshift space is a useful probe of cosmology. 
This geometrical test of dark energy will complement and extend currently used geometrical probes such as measuring  the light curves of Type Ia supernovae (SN)
\citep{Riess:1998cb,Perlmutter:1998np, 2011ApJ...730..119R} and applications of the Alcock-Paczynski test to baryonic acoustic oscillations 
\citep[e.g.][]{2011arXiv1108.2637B}.

In a Friedmann-Robertson-Walker universe, pairs of galaxies should be distributed
with random orientations if the fundamental assumptions of homogeneity
and isotropy are correct. This simple test of cosmology is complicated by two effects:
firstly, we do not observe galaxies directly in real space but in redshift space, where
peculiar velocities, distinct from the Hubble flow, displace the  position of
a galaxy along the line of sight from its true position. This introduces a preferred direction, with the
result that galaxy pairs are no longer randomly distributed. Secondly, in order to convert
observed angles and redshifts into comoving distances, an observer needs to assume a
cosmological model. An intrinsically  spherical object, such as a cluster of galaxies,
or a spherically symmetrical distribution, such as the distribution of galaxy pairs we
consider in this paper, will appear distorted if measured assuming a cosmology that does
not match the true underlying cosmology of the Universe \citep{1979Natur.281..358A}.

Based on this idea, \citet{1994MNRAS.269.1077P} proposed a test where the hypothetical
sphere proposed by  \citet{1979Natur.281..358A} is replaced by randomly orientated quasar pairs.
In the absence of peculiar motions, a large sample of quasar pairs should have a uniform
distribution in the cosine of the angle between each pair, if the correct cosmology is adopted.
\citet{2010Natur.468..539M} developed this test by modelling the effect of the redshift
space distortions as a Doppler shift in the positions of the galaxies. They applied this model  to
galaxy pairs in the Sloan Digital Sky Survey (SDSS) \citep{2009ApJS..182..543A} at low redshift and
the DEEP2 galaxy redshift survey \citep{2004ApJ...609..525C} at $z\sim 1.3$.
After selecting galaxy pairs according to a set of constraints discussed in Section 4,
Marinoni \& Buzzi  were left with a sample of 721 pairs in the SDSS DR7 at $z \sim 0$. Mainly due to the small sample size, Marinoni \& Buzzi were only able to distinguish
a $\Lambda$CDM cosmology  from somewhat extreme alternatives, namely an Einstein-de Sitter Universe and an open
universe with no dark energy. We note that these models have already been ruled out by other tests.
 This does not, however, imply that the test cannot
be used to yield competitive constraints on dark energy with a larger sample of pairs.

Note that the Alcock-Paczynski test measures a distortion parameter which is proportional to the  angular diameter distance at the redshift of the 
object, $D_A(z)$ multiplied by the Hubble rate, $H(z)$. The test we propose in this paper models the distribution of galaxy pairs in real and redshift space assuming a distant observer
approximation. This assumption removes any dependence on the angular diameter distance in the distortion parameter. As a result this technique allows us to 
measure the Hubble rate directly.

In this paper, using subhalo pairs in large volume N-body simulations, we test the method of Marinoni
\& Buzzi and its potential to distinguish between cosmologies. First we focus on the selection criteria
necessary to provide a homogeneous sample of pairs whose distribution in redshift space agrees with
the theoretical model of Marinoni \& Buzzi. With robust selection criteria, we then apply
this test to different simulations to see if these dark energy models can be distinguished from $\Lambda$CDM.
A critical assumption made in the analysis by Marinoni \& Buzzi is that the
normalization of a theoretical model of the pair distribution does not evolve with redshift.
We show, using numerical simulations, that this assumption is incorrect.
We also consider the practical difficulties associated with obtaining an
accurate measurement of this normalization parameter  observationally. We demonstrate that the test,
as originally proposed, suffers from large systematics which limit its utility. We present an improved
methodology which uses N-body simulations and does not require prior knowledge of the true cosmological model.

Most attention to date has focused on cosmological tests which require
measurements on large scales, such as the rate at which cosmic structures
grow \citep{Guzzo:2008ac,Wang:2007ht}, the apparent location of
baryonic acoustic oscillations \citep{2009MNRAS.400.1643S,2010MNRAS.tmp..776B}
and the projected matter density as measured through weak lensing \citep{2007Natur.445..286M}.
It is important to expand this arsenal of tests. This introduces sensitivity to different systematics,
which, alongside results from other probes, will lead ultimately to a more convincing
measurement of the properties of dark energy. Also, it is useful to devise new tests 
which are not reliant on measuring the galaxy distribution on the very largest scales, 
thereby avoiding the need for an accurate determination of the mean galaxy density
\citep[for another example see][]{2011arXiv1106.6145N}. The test proposed 
in this paper requires a large volume simply to accumulate a large sample 
of galaxy pairs; there is no requirement implied on the accuracy of the 
photometry across a survey used for this purpose. 

A novel feature of our analysis is the use of N-body simulations to validate and improve upon the
methodology proposed by Marinoni \& Buzzi. Recent work has shown that numerical simulations of structure formation have an
important role to play in modelling cosmological probes and assessing potential systematic errors.
Angulo et~al. (2008) demonstrated that the shape of the power spectrum of galaxy clustering is substantially
different from the predictions of linear perturbation theory even on very large scales
\citep[see also][]{ 2007PhRvD..75f3512S,2008PhRvD..77d3525S,2008ApJ...686...13S,2010MNRAS.401.2181J}.
The simulation results led to revised analyses of baryonic acoustic oscillations,
which either attempt to model the distortions introduced by nonlinearities and redshift
space, or to reconstruct the linear theory signal \citep{2007ApJ...665...14S,
2008MNRAS.390.1470S, 2010MNRAS.408.2397M,2011arXiv1107.4097M}. Similarly, N-body simulations
have demonstrated that the measurement of the growth factor from redshift space distortions
requires careful modelling \citep{2011MNRAS.410.2081J,2011ApJ...726....5O}. \citet{2011ApJ...727L...9J}
showed that a na\"{i}ve application of a linear theory model for the distortion of clustering in
redshift space can lead to a catastrophic misinterpretation of the measured growth factor.
The study in this paper is in a similar spirit; the availability of N-body simulations to model 
the pair distribution  allows us to devise an improved cosmological probe.

The outline of this paper is as follows: In Section 2, we review the theoretical model of
\citet{2010Natur.468..539M} for the anisotropic distribution of pairs and its dependence on cosmology.
In Section 3 we discuss the quintessence dark energy models and the N-body simulations used in this paper.
In Section 4 we list and test the selection criteria used to select a homogeneous sample
of galaxy pairs from the N-body simulations whose distribution agrees with the theoretical predictions.
In Section 5 we present our results, comparing the theory with measurements from simulations, for the two dark
energy and the $\Lambda$CDM cosmologies and demonstrate that a robust test of cosmology can
only be achieved by combining observations with numerical simulations.
In Section 6 and 7 we present our summary and conclusions.

\section{Theoretical background: the distribution of galaxy pairs \label{sect}}

\begin{figure}
\center
{\epsfxsize=3.truecm
\epsfbox[47 483 149 685]{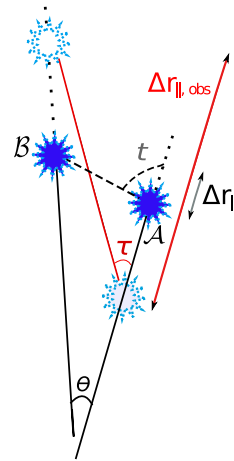}}
\caption{An illustration of the actual, $t$ and observed, $\tau$, angles subtended between a pair of galaxies, $A$ and $B$, and the line of sight through $A$ in real space (black) and in redshift space (red) respectively. At the 
observer's position, the pair subtend an angle $\theta$. The comoving line of sight separation of the pair 
is shown as $\Delta r_{\parallel}$ and $\Delta r_{\parallel, \tiny \mbox{obs}}$ in real and redshift space respectively. 
\label{cartoon}}
\end{figure}

Following the derivation and discussion in \citet{2010Natur.468..539M} \citep[see also the alternative derivation in][]{1994MNRAS.269.1077P},
 let us consider a pair of gravitationally bound galaxies, $A$ and $B$, at an observed angular separation, $\theta$, as shown in Fig. \ref{cartoon}.
In a flat universe, the 
angle that galaxy $B$ subtends at galaxy $A$, as measured from the observer's line of sight through $A$, which we refer to as the tilt angle,
 $t$, 
can be written as 
\begin{eqnarray}
\sin^2 t &=& [1+ \left( \cot \theta  - \frac{\chi_A }{\chi_B \sin \theta}\right)^2 ]^{-1} \, ,
\label{sin2t}
\end{eqnarray}
(see \citet{2010Natur.468..539M} for the general expression in a curved universe). Here $\chi_{A (B)}$ is the radial comoving distance to galaxy $A (B)$ which is given by
\begin{eqnarray}
\chi(z) &=& \frac{c}{a_0} \int^z_0 \frac{{\rm{d}}z'}{H(z')} \, ,
\label{chi}
\end{eqnarray}
where  $a = 1/(1+z)$ is the scale factor of the universe with current value $a_0=1$, $c$ is the speed of light and $H = \dot{a}/a$ is the Hubble
parameter with current value $H_0 = 71.5 $ km/s/Mpc \citep{2009MNRAS.400.1643S}.
In an isotropic and homogeneous universe, the orientation of pairs of bound galaxies will be randomly distributed, allowing us to predict the probability
distribution for $t$, $F(t)$, and a measure of the distribution $\mu = \langle \sin^2 t \rangle$.
In the absence of peculiar velocities, an observer calculating the ensemble average of Eq. \ref{sin2t}  should find
a value of
$\mu = 2/3$, independent of cosmology, as long as the correct cosmology is assumed when converting angles and redshifts to comoving distances for each member of the pair.

Galaxies have velocities in addition to the Hubble flow which result in inferred positions which appear displaced along the line of sight. 
As a result the true angle $t$ between a pair of gravitationally bound galaxies 
will appear as an angle, $\tau$ (see Fig. \ref{cartoon}).
The angle subtended between the pair of galaxies in redshift space calculated according to Eq. \ref{sin2t}  corresponds to a measurement of the angle
 $\tau$ and not $t$, and the average over all pairs will no longer be a random distribution.
The result is a skewed distribution whose mean will differ from the expected value for a uniform distribution of $\mu = 2/3$. 
Adopting the correct cosmology
 to calculate the ensemble average in Eq. \ref{sin2t} will then provide a measure of the mean of the apparent distribution in redshift space after intrinsic peculiar velocities distort the orientation of the pairs.
Marinoni \& Buzzi modelled this distortion as a simple
 Doppler shift where the observed line of sight separation is related to the actual separation, to first order in $v/c$, by
\begin{eqnarray}
{\rm{d}}r_{\mbox{\tiny obs}} &=& {\rm{d}}r + \frac{{\rm{d}}v_{\parallel}}{H(z)}(1 +z) \, ,
\end{eqnarray}
 where ${\rm{d}}r_{\mbox{\tiny obs}}$ and ${\rm{d}}r$ are the observed and actual line of sight separations of a pair of galaxies $A$ and $B$ and ${\rm{d}}v_{\parallel} = 
v_A \cdot \hat{r}_A -v_B \cdot \hat{r}_B$ is the line of sight peculiar velocity difference, where $\hat{r}_{A (B)}$ represent unit vectors in the direction of each galaxy in the pair. 
The line of sight comoving separation is ${\rm {d}} r = {\rm d}\chi$, as given in Eq. \ref{chi}. 
Note this equation is a
result of relating the position observed in redshift space to the
actual position in real space as ${\rm{d}} z_{\mbox{\tiny obs}} = {\rm{d}}z + {\rm{d}}v_{\parallel}$ 
e.g \citet{1998ASSL..231..185H}.
 In the distant observer approximation, the separation between galaxies is assumed to be 
small compared to the distance between them and the observer. Under this assumption 
the observed comoving transverse separation is equal to the true transverse separation of the pair, ${\rm{d}}r_{\perp ,\mbox{\tiny obs}} \approx {\rm{d}}r_{\perp}$. 
If the redshift difference of the pair $\Delta z$ is a lot less than unity such that $\Delta z \approx {\rm d}z$, then
the 
observed, $\tau$, and actual tilt, $t$, of the pair can be simply related by the observed, 
$ \Delta r_{\parallel, \mbox{\tiny obs}}$ and the actual line of sight finite separation, $ \Delta r_{\parallel}$,  according to
\begin{eqnarray}
\frac{\tan t}{\tan \tau} &=& \frac{\Delta r_{\parallel , \mbox{ \tiny obs}}}{\Delta r_{\parallel}} 
= 1 + \frac{(1+z)}{H(z)}\frac{\Delta v_{\parallel}}{\Delta r_{\parallel}} \, .
\label{tantan}
\end{eqnarray}
The relation given in Eq. \ref{tantan} can then be used to transform the true distribution of galaxy pairs, $F(t)$, into the apparent distribution, $\Psi(\tau)$. 
Using conservation of probability, Marinoni \& Buzzi derived the probability distribution function of the apparent angle written in terms of the true angle as 
\begin{eqnarray}
\Psi(\tau) {\rm{d}}\tau &=& F(t){\rm{d}}t \,. 
\end{eqnarray} 
From this it follows that $\Psi(\tau) $ is given by
\begin{eqnarray}
\Psi(\tau){\rm{d}}\tau &=& \frac{1}{2}\frac{(1+\sigma^2)(1+ \tan^2 \tau)}{\left( 1 + (1+\sigma^2)\tan^2 \tau \right)^{3/2}}|\tan \tau| {\rm{d}} \tau \, ,
\label{dist}
\end{eqnarray}
and the parameter $\sigma$ depends on the cosmological expansion history as
\begin{eqnarray}
\sigma^2 ( z, \Omega) =  2 \left \langle \frac{\Delta v_{\parallel}}{\Delta r } \right \rangle \frac{1+z}{H(z)} +  \alpha^2  \frac{H^2_0(1+z)^2}{H^2(z)}  \, .
\label{sig}
\end{eqnarray}
The normalization parameter $\alpha$ is given by
\begin{eqnarray}
\alpha &=& H^{-1}_0 \left(\left \langle \frac{\Delta v^2_{\parallel}}{\Delta r^2} \right \rangle\right)^{1/2} \,.
\end{eqnarray} 
The first moment of the distribution $\Psi(\tau)$, referred to by Marinoni \& Buzzi as the \lq average anisotropy of pairs\rq \,( the AAP function from now on), is given by 
\begin{eqnarray}
\mu_{\mbox{\tiny obs}} &=& \frac{(1+\sigma^2)\,\mbox{arctan}(\sigma) - \sigma}{\sigma^3} \, .
\label{aapfunction}
\end{eqnarray}
In Eq. \ref{aapfunction}, the parameter $\sigma$ depends on the expansion history in a particular cosmological model, $H(z)$, as given in Eq. \ref{sig}.
 Marinoni \& Buzzi set the first term on the right hand side of Eq. \ref{sig} to zero on the  assumption that the comoving separation of pairs and their radial peculiar velocities are uncorrelated, 
$\langle \Delta v_{\parallel}/\Delta r\rangle=0$. We shall discuss this assumption further in Section 4.

The original Alcock-Paczynski test, when applied to a spherical
object, measures a distortion parameter, the ratio of the tangential
and radial distances, which is proportional to $D_A(z) H(z)$ and is
unity if the correct cosmological model is assumed and there are no
redshift space effects i.e. there is no distortion of the spherical
object.
The Alcock-Paczynski test applied in this paper compares the distribution of pair
angles in real and redshift space in the distant observer
approximation,  ${\rm{d}}r_{\perp ,\mbox{\tiny obs}} \approx {\rm{d}}r_{\perp}$,
which gives rise to a distortion parameter which is independent of $D_A$.
The distortion is estimated, after modelling redshift space
effects,
by comparing the distribution of the angles $t$ and $\tau$, and only
depends on $H(z)$.

 Using pairs of galaxies in a survey, an observer can measure the average orientation using Eq. \ref{sin2t} which should be equal to 
the AAP function in  Eq. \ref{aapfunction} if the correct cosmology is assumed and the observer is able to measure $\alpha$ precisely in order to 
fully specify $\Psi(\tau)$. 
In this paper we perform this exact test using pairs of subhaloes in N-body simulations of different cosmologies.
In practise in a galaxy survey the parameter $\alpha$ can  be determined in two ways: firstly, at low redshifts, where the peculiar velocities of the pair can be measured
by combining a redshift independent distance measurement, e.g. luminosity distances from Type Ia supernovae, the Tully-Fisher relation or the $D_n -  \sigma$ relation
 \citep[see e.g.][]{2000ApJ...544..636C, 2000AJ....120...95D,2000AJ....119..102B,2000ASPC..201..254B}, with
the measured redshift of the galaxy. 
The uncertainties associated with the redshift independent luminosity distance measurements are large, $\sim 10-20$\% for the Tully-Fisher or $D_n - \sigma$ relations,
and $\sim 5-10$\% for supernovae or the surface brightness fluctuation method \citep{2001astro.ph.10344B, 2001ApJ...546..681T}. These uncertainties on the redshift independent
distance measurements propagate into larger errors for the peculiar velocities,
making it almost impossible to accurately measure the peculiar velocity of a single galaxy.
 The second method to determine $\alpha$ observationally, which we shall assess in this paper after considering the ideal case of measuring $\alpha$ from the simulations 
using Eq. 8, is to fit  to the
measured distribution of pairs at each redshift using  Eq. \ref{dist}.

One of the key assumptions made by Marinoni \& Buzzi 
 is that the normalization
factor $\alpha$ is constant for all redshifts and for different galaxy selections. At $z \approx 0$, Marinoni \& Buzzi obtained 
$\alpha = 5.79^{+0.32}_{-0.35}$,   using binaries in
 the SDSS \citep{2009ApJS..182..543A}. 
Marinoni \& Buzzi  obtained this value  by fitting Eq. \ref{dist} to the observed distribution at 
$z \approx 0$.
We explicitly test this assumption in this paper where it is possible to measure $\alpha$ directly from the N-body simulations at each redshift. We can also compare the 
predictions
of the AAP function using the best fit value for $\alpha$ obtained at $z=0$, 
instead of normalizing the function at each redshift. This will allow us to see if the value of $\alpha$ really is independent of redshift.

\section{Trial samples of pairs from numerical simulations}

As a test of the method proposed by Marinoni \& Buzzi which was  outlined in Section 2, we apply it to different cosmologies, focussing on quintessence models.
In Section \ref{sub20} we discuss the two quintessence dark energy models we take as examples and highlight the main differences between these  and the concordance
cosmological model.
In Section \ref{sim21} we describe the simulations carried out.

\subsection{Quintessence dark energy \label{sub20}}

Numerous quintessence dark energy models have been considered as an alternative to the concordance cosmology \citep[see e.g.][]{Ratra:1987rm, Ferreira:1997hj, Copeland:2006wr, 2008MPLA...23.1252M}.
We focus on two interesting examples which are representative of a wider class of quintessence models, scalar fields which evolve in time, which are viable alternative cosmologies.

One of the models we consider has substantial differences to $\Lambda$CDM and can be considered as an \lq early\rq \, dark energy model which features non-negligible amounts of dark energy at high redshifts. 
This quintessence dark energy model features an exponential term in the scalar field potential  which pushes the  dark energy equation of state to $w_0 =-0.82$ today  \citep{Brax:1999gp}.
We refer to this model as
the SUGRA model.
The second quintessence dark energy model, which we refer to as INV, has been shown to produce a similar expansion history and non-linear growth of structure to those in a $\Lambda$CDM cosmology 
\citep{2010MNRAS.401.2181J, 2011MNRAS.410.2081J} 
and will provide a measure of the sensitivity of 
the test we perform in this paper.
The INV model has an inverse power law potential $V(\phi) = \Lambda^{\beta +4}/\phi$ for the scalar field $\phi$ \citep{Zlatev:1998tr}. The values
of the constants $\Lambda$ and $\beta$ are fixed by the current value of the dark energy density \citep[see e.g.][]{Corasaniti:2002vg}. 

The dark energy equation of state for these quintessence models can be accurately described over a wide range of redshifts using  four parameters  \citep{Corasaniti:2002vg}.
The variables used are: $w_0$, the current dark energy equation of state; $w_{\rm{m}}$,
 the value of $w$ during the matter dominated era; $a_{\rm{m}}$, the scale factor at which the dark
energy equation of state changes from its value during the matter dominated era to its
present value, and $\Delta_{\rm{m}}$, the width of the transition in the expansion factor. 
For the SUGRA model these parameters are $w_{0} = -0.82 , w_{\rm{m}} =
-0.18, a_{\rm{m}} = 0.1$ and $\Delta_{\rm{m}} = 0.7$.
For the INV model the values of the parameters are $w_{0} = -0.79 , w_{\rm{m}} = 
-0.67, a_{\rm{m}} = 0.29$ and $\Delta_{\rm{m}} = 0.4$ \citep{2010MNRAS.401.2181J}.

The dark energy models have different expansion histories to $\Lambda$CDM and so when compared to the currently available observations may favour
 different best fitting 
values of the cosmological parameters
 \citep[see][for a discussion]{2010MNRAS.401.2181J}.
As our starting point, we consider a $\Lambda$CDM model with
the following cosmological parameters:
$\Omega_{\rm m} = 0.26$,
 $\Omega_{\rm{DE}}=0.74$, $\Omega_{\rm b} = 0.044$,
$h = 0.715$, where $H_0 = 100h$ km/s/Mpc and a spectral tilt of $n_{\mbox{s}} =0.96$ \citep{2009MNRAS.400.1643S}.
The  linear theory rms fluctuation
in spheres of radius 8 $h^{-1}$ Mpc is set to be  $\sigma_8 = 0.8$.
In the simulations discussed in this paper, the $\Lambda$CDM values for $\Omega_{\rm m}$ and $H_0$  were used for the INV dark energy model while for the SUGRA model the best fit parameters used were
$\Omega_{\rm m} = 0.243$ and  $H_0 = 67.73$km/s/Mpc \citep[see][ for more details]{2010MNRAS.401.2181J}.
Both of these models are  consistent with current observations of Supernovae  Type Ia lightcurves \citep{Kowalski:2008ez}, baryonic acoustic oscillations \citep{Percival:2007yw, 2009MNRAS.400.1643S} and the WMAP 7 year measurements of the
cosmic microwave background \citep{2010arXiv1001.4538K}.
A detailed study of both of these models compared to a $\Lambda$CDM cosmology can be found in \citet{2010MNRAS.401.2181J}.

Note \citet{2010MNRAS.401.2181J} showed that the INV model was indistinguishable from $\Lambda$CDM for several cosmological probes such as measurements of the 
halo mass function, BAO peak positions and growth factor. This model provides us with a significant test of the discriminatory power of the technique proposed
in this paper.

\begin{figure}
\center
{\epsfxsize=7.5truecm
\epsfbox[70 364 308 611]{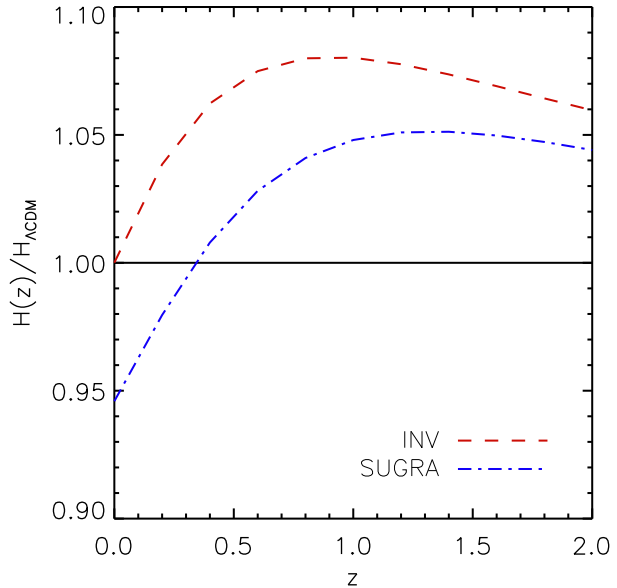}}
\caption{ The ratio of the Hubble rate in the two quintessence dark energy models simulated  to that in a $\Lambda$CDM cosmology plotted as a function of redshift. Note the SUGRA quintessence model has a current value for the Hubble parameter of $H_0=67.73
$km/s/Mpc,
consistent with observations (see Section 3.1 for details).
\label{H}}
\end{figure}

The ratio of the Hubble parameter in each quintessence dark energy model to that in a $\Lambda$CDM cosmology  is shown as a function of redshift 
in Fig. \ref{H}.

\subsection{N-body simulations  \label{sim21}}

The simulations were carried out at the Institute of Computational Cosmology using a memory efficient version of the  TreePM
  code  {\tt Gadget-2}, called {\tt L-Gadget-2} \citep{Springel:2005mi}.
The simulations use $N=646^3 \sim 269 \times 10^6$ particles to represent the dark matter in a  computational box of
comoving length $1500 h^{-1}$Mpc. We shall refer to these simulations as the low resolution runs in Section \ref{sub3.1}.
 We chose a  comoving softening length of  $\epsilon = 50 h^{-1}$kpc.
The particle mass in the simulation is $9.02 \times 10^{11}  h^{-1}
M_{\sun}$ with a mean interparticle separation of
$r \sim 2.3$ $h^{-1}$Mpc.
We also consider a higher resolution simulation of the $\Lambda$CDM cosmology with the same boxsize as above but with 1024$^3$ dark matter particles, approximately four times the 
number of particles used in the lower resolution simulation.

The initial conditions of the particle load were set up with a
 glass configuration of particles \citep{1994RvMA....7..255W,Baugh:1995hv}.
The particles are perturbed from the glass using the Zel'dovich approximation which
can induce small scale transients in the measured power spectrum. These transients die away after $\sim 3-10$ expansion factors from the starting redshift
\citep{Baugh:1995hv, Smith:2002dz}.
In order to limit the effects of the initial displacement scheme
 we chose a starting redshift of $z=200$.
The linear theory power spectrum used to generate the initial
conditions was created using the CAMB package of \citet{Lewis:2002ah}. The linear theory power spectrum for the two dark energy models was computed using the 
Parameterized Post-Friedmann (PPF) module \citep{Fang:2008sn} for CAMB which takes into account the effects 
of a dynamical dark energy equation of state and dark energy perturbations \citep{2010MNRAS.401.2181J}.

\begin{figure}
{\epsfxsize=8.truecm
\epsfbox[71 360 332 612]{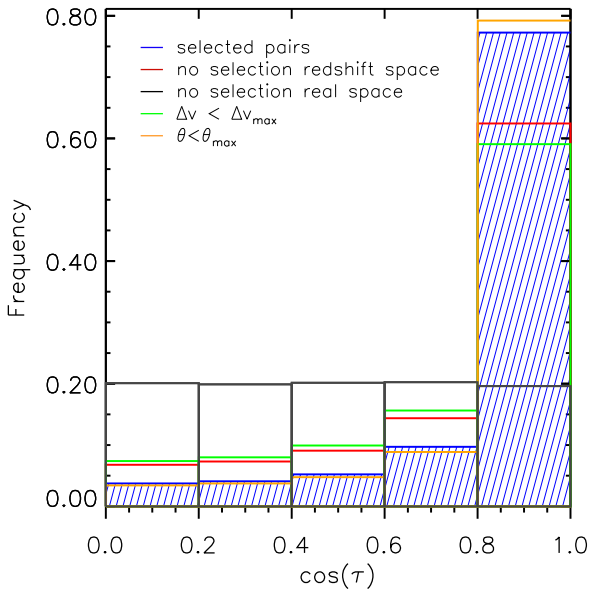}}
\caption{ The distribution of the cosine of the apparent inclination angle, $\cos(\tau)$, of pairs of subhaloes in a $\Lambda$CDM simulation at $z=0$.
The distribution of $\cos(\tau)$ for all the subhalo pairs in redshift (real) space is shown in red (black).
Subhalo pairs which are selected according to the criteria discussed in Section 4 are shown as blue hashed boxes. 
Selecting pairs with only a cut in $\theta_{\mbox{\tiny max}}$ gives rise to the distribution shown in orange. Selecting pairs with only a cut in $\Delta v_{\mbox{\tiny max}}$ gives rise to the distribution shown in green.
}
\label{pdf}
\end{figure}

Dark matter haloes were identified in the simulation outputs using a friends-of-friends (FOF) percolation algorithm with a linking length of $b=0.2$ times the 
mean interparticle separation \citep{1985ApJ...292..371D}. The {\sc{SUBFIND}} algorithm \citep{2001MNRAS.328..726S} was then run on these 
halo catalogues to identify  self bound subhaloes
at each redshift.
Note that the subhaloes are not necessarily bound to the main FOF halo.
In this paper, pairs of subhaloes within a common FOF halo are used as a proxy for pairs of galaxies. 

The minimum number of particles per halo and subhalo is 10 and we select only haloes that have at least two subhaloes (i.e. a minimum of 20 particles in the FOF 
group). In Fig. 6 we show that our results are not affected by our choice of minimum FOF halo mass.

The position of each subhalo in redshift space is 
computed by perturbing its comoving position in real space using the line of sight centre of mass 
velocity of the subhalo relative to an observer placed at the origin of the box. 
At a given redshift $\tilde{z}>0$ the observer is still assumed to be at the origin of the box at $z=0$ which requires us to add the comoving distance from $z=0$ to $\tilde{z}$ to the subhalo positions.
In Section 4 we discuss the selection criterion for subhalo pairs in relation to the radius $R_{200}$ of the parent halo where the mean density is 200 times the critical value.

\begin{figure*}
{\epsfxsize=16.truecm
\epsfbox[69 365 545 616]{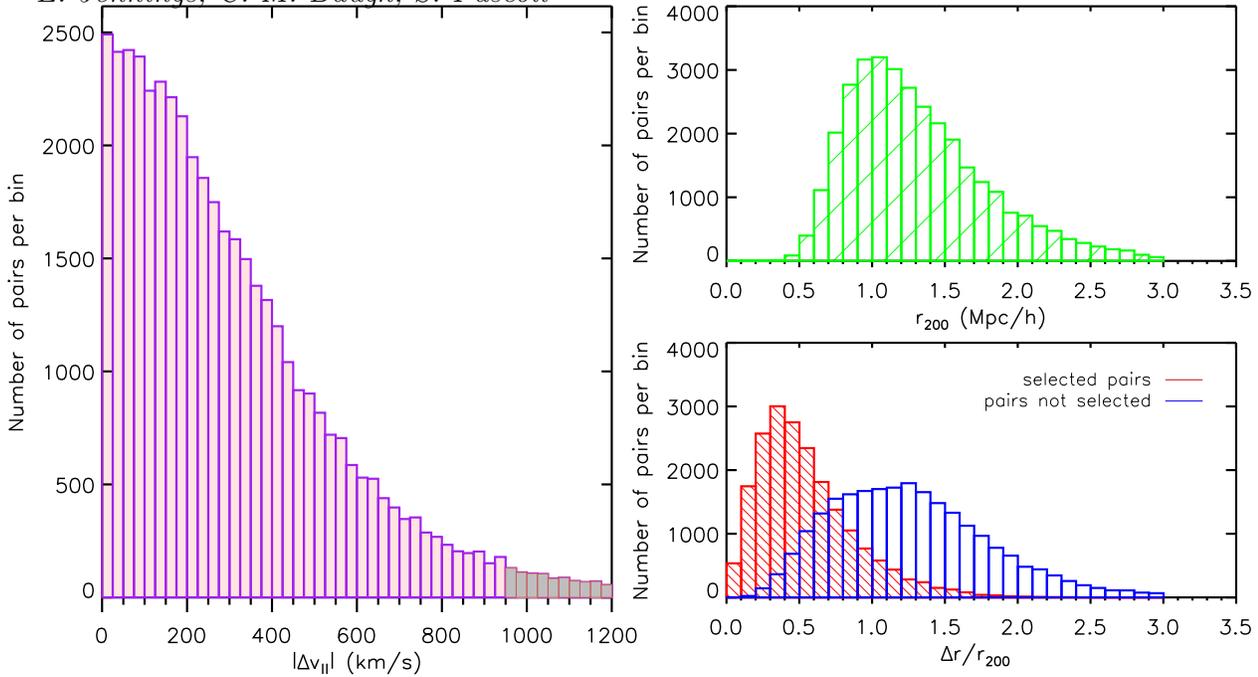}}
\caption{ Left panel: The distribution of the line of sight peculiar velocity difference, $\Delta v_{\parallel}$, for all pairs of subhaloes in the lower resolution $\Lambda$CDM simulation at $z=0$. Pairs to the left of the 
 grey shaded region represent those subhaloes that have been selected (95\% of 
the distribution).
Lower right panel: The comoving transverse separation in redshift space, $\Delta r_{\perp}$, of pairs of subhaloes in the lower resolution $\Lambda$CDM simulation at 
$z=0$ plotted as a fraction of $R_{200}$. The selected subhaloes are shown as a red hashed region while those not selected are shown in blue. 
Upper right panel: The distribution of the radius, $R_{200}$, for each parent halo is shown as a green hashed region.
\label{dvdr}}
\end{figure*}

\section{Calibrating the  method  using simulations \label{pair}}


From the N-body simulations we know which pairs of subhaloes are in the same FOF halo. However, this does not guarantee that these
objects are gravitationally bound to the FOF halo. There are too many haloes in our simulations to check explicitly for binding, so we will use proxies instead.
This will allow us to make contact with the observational selection applied by Marinoni \& Buzzi and to see how 
their cuts translate into cuts in simulation quantities.
We investigate these selection criteria and provide robust selection cuts which are independent of cosmology and redshift.

In particular we address the following question: how do we construct a sample of pairs that matches the theoretical expectation for the AAP function, in
the most favourable case in which  we know the correct cosmology?
Using information output from the  simulations about the subhaloes selected and the properties of the parent halo, (e.g. the FOF algorithm
returns $R_{200}$),  we can quantify the definition of a close pair in a rigorous way.
If we selected only bound pairs we would expect good agreement with the AAP function, provided that we know
the correct cosmological model. As a first approach to identify suitable pairs, we shall select subhaloes which are within $R_{200}$ of the main halo, i.e. $\Delta r_{\perp,\mbox{\tiny max}} = R_{200}$, with no other restrictions on velocity or
distance to a nearest neighbour.
As this information is not available to an observer, our second approach will be to translate these selection criteria into observable quantities such as the angle $\theta$ between a pair of subhaloes.

 Marinoni \& Buzzi  used the following selection criteria to pick their sample of pairs: (1)
a maximum line of sight velocity difference of the pair $\Delta V  = 700$ km/s to avoid projection of neighbouring systems, 
(2) a maximum comoving transverse separation of $\Delta r_{\perp,\mbox{\tiny max}} = 0.7$ Mpc$/h$,   
(3) a minimum comoving transverse separation $\Delta r_{\perp ,\mbox{\tiny min}} = 20$ kpc$/h$ 
and  (4) a minimum comoving distance from the centre of the galaxy pair to  another galaxy.
 The latter two conditions avoid selecting pairs which may be in the process of merging or which are interacting with another galaxy.
 The value for the
maximum velocity difference was chosen such that the relative increase $\Delta N/N $ in the sample size was $<1\%$ when the velocity cut was increased by 100 km/s,
 while the maximum comoving transverse 
separation was chosen to be
equal to the distance from Andromeda to the Milky Way.

 Fig. \ref{pdf} shows the measured distributions of the orientation of subhalo pairs in real and redshift space in the low resolution $\Lambda$CDM simulation
at $z=0$.
The starting point is the sample of subhalo pairs within a common FOF halo, without any further selection. This is shown in real space by the black histogram in Fig. \ref{pdf}.
Note that for the lower resolution $\Lambda$CDM simulation there are 
approximately 65,000 subhalo pairs at $z=0$.
The real space distribution of the tilt follows the expected random distribution and is uniform in $\cos(\tau)$.
The distribution of all subhalo pairs in redshift space is shown in red which is clearly skewed. The mean of this distribution differs from the prediction of the AAP function by $\sim 40$\%.
Applying the final set of cuts to this overall sample as outlined below leaves approximately 19,000 pairs, and 
produces the blue hashed region which is skewed towards smaller angles and agrees with the predictions of the AAP function given in Eq. \ref{aapfunction} to within 0.5\%. We discuss the selection cuts that give rise to this blue hashed region 
below.
A comparison of the red and blue histograms in Fig. \ref{pdf} shows that in redshift space if no selection cuts are made to isolate bound pairs the distribution is clearly randomized by
outliers.

In an attempt to isolate subhaloes that are gravitationally bound to their parent FOF halo and hence 
 to test if their orientations in redshift space are distributed according to the predictions of the AAP function,
 we first select pairs of subhaloes within $R_{200}$ and exclude all other pairs.
We find that this sample of pairs has a non-negligible correlation between $\Delta v_{\parallel}$ and $\Delta r$ such that $\langle \Delta v_{\parallel}/ \Delta r \rangle \ne 0 $.
As a result,  we use the full expression in Eq. \ref{sig} for the parameter $\sigma$. 
This gives an AAP function which is in remarkably good agreement with the measured mean of the distribution, the ensemble average of 
Eq. 1, at $z=0$ in
 a $\Lambda$CDM simulation, to better than a percent.  This agreement diminishes at higher redshifts, 
 with the AAP function and the measured mean differing by 10-20\% over the redshift range $z=0.25 - 1$. 

It is possible to remove  subhalo pairs which have   $\langle \Delta v_{\parallel}/ \Delta r \rangle \ne 0 $ by selecting    pairs according to
an upper limit in the line of sight peculiar velocity difference, $\Delta v_{\tiny \mbox{max}}$. 
The velocity difference of pairs of galaxies is related to the common gravitational potential of the pair which, in most cases, is weakly correlated with their separation. However, we find that pairs with large velocity differences 
have non-zero correlations, e.g. in the $\Lambda$CDM simulation at $z=0$, using all subhalo pairs with $\Delta v > 950$km/s we find $\langle \Delta v_{\parallel}/ \Delta r \rangle =8.5h$ km/s/Mpc. 
Observationally these subhaloes would not be detected as the apparent tilt between the pair is approximately zero, due to their
 large peculiar velocity difference, and as a result the pair would lie along the same line 
of sight. 
In Fig. \ref{dvdr} we plot the distribution of the line of sight peculiar velocity difference $\Delta v$
in the left panel, for all subhaloes in the lower resolution
$\Lambda$CDM  simulation. The grey shaded region corresponds to the selection cut in $ \Delta v$.
Once we remove any correlated pairs from the sample, and impose the restriction that $\Delta r_{\perp,\mbox{\tiny max}} = R_{200}$, we find that the measured mean and the predicted AAP function agree extremely well in the 
redshift range $z=0-2$. We present these results in more detail in the following section. Note the first term in the expression for $\sigma$, Eq. \ref{sig}, is now negligible as we have removed any correlated pairs.

As $R_{200}$
is not an observable quantity, the next step is to see if this cut can be translated into a cut in $\theta$, the observed angular separation of the pair.
In Fig. \ref{dvdr} we plot the distribution of 
the comoving transverse separation $\Delta r_{\perp}$ as a fraction of $R_{200}$ in the lower right panel, for all subhaloes in the lower resolution
$\Lambda$CDM  simulation. 
The comoving transverse separation $\Delta r_{\perp}$ as a fraction of $R_{200}$ for the subhaloes that are selected
 by $\theta < \theta_{\mbox{\tiny max}}$
 is shown as a
red hashed region while the distribution of those not selected is shown in blue.
In the upper right panel in Fig. \ref{dvdr} the distribution of $R_{200}$ for the parent haloes is plotted in green.
Selecting pairs according to $\theta_{\mbox{\tiny max}}$  gives rise to a sample containing most of the subhalo pairs which have $\Delta r_{\perp} <R_{200}$, with only
a small number of pairs with $\Delta r_{\perp} >R_{200}$ that happen to lie at an angle
 $\theta < \theta_{\mbox{\tiny max}}$.

We find the following selection rules give rise to a population of pairs whose measured moment matches the predictions of
the AAP function extremely well, provided the correct cosmological model is assumed (see Section \ref{sub3.1}):
\begin{itemize}
\item The upper limit of the line of sight velocity difference should correspond
to retaining 95\% of the total distribution of pairs in the sample (grey shaded region in the left panel of Fig \ref{dvdr}). 
\item The maximum observed separation of a pair, $\theta_{\mbox{\tiny max}}$, should correspond to retaining $50-60$\% of the distribution for all pairs.
\end{itemize}

Subhalo pairs in redshift space  chosen according to the two selection criterion given above give rise to the blue hashed region  shown
in Fig. \ref{pdf}.
For this $\Lambda$CDM simulation at $z=0$ this corresponds to 
approximately 19,000 pairs with $\Delta v <950$ km/s (95\% of the distribution) and
$\theta < 6.5 \times 10^{-4}$ rad (50\% of the distribution).
{bf
Note these specific values quoted for $\Delta v$ and $\theta$ are only for illustration. The selection criteria presented in the two bullet points above should
be applied the to parent sample of galaxy pairs when implementing this test.
}
If we select pairs by restricting $\Delta v$ only, we retain 38,000 subhalo pairs and then the difference between the measured mean and the corresponding 
AAP function is approximately 30\%. 
This distribution is shown in green
in Fig. \ref{pdf}. Selecting pairs with $\theta <\theta_{\mbox{\tiny max}}$ and no restriction on $\Delta v$, i.e. including correlated pairs with
$\langle \Delta v_{\parallel}/ \Delta r \rangle \ne 0 $,
  results in a mean that differs from the corresponding AAP function by less than 1\%, provided the full expression for $\sigma$ in Eq. 1 is used  
(shown in orange
in Fig. \ref{pdf}). Note if the full expression is not used then the difference is 4\%.
We find that the measured moment is most sensitive to the first two selection criterion  chosen by  Marinoni \& Buzzi  and relatively insensitive to the minimum comoving separation of the pair and the comoving distance to the nearest neighbour. Note this is partly 
because we only consider pairs of subhalos from the same halo.

\begin{figure}
{\epsfxsize=8.5truecm
\epsfbox[56 366 366 650]{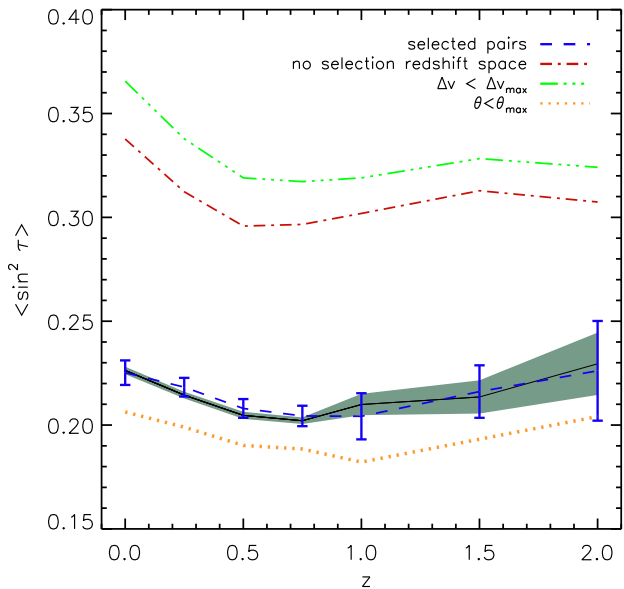}}
\caption{ The sample means of the anisotropic distributions  of pair tilt angles shown in Fig. \ref{pdf} for subhaloes  in the low resolution $\Lambda$CDM simulation,
as a function of redshift. The mean of the distribution of all pairs in redshift space (no selection cuts) is shown as a red dot-dashed line. 
The measured mean for pairs selected with only a cut in $\Delta v$ ($\theta$) is 
shown as a green dot-dot-dashed  (orange dotted) line.
Once we impose a cut in both $\Delta v$ and $\theta$, the measured mean of the distribution (blue dashed line) agrees with the corresponding predicted AAP function 
(solid black line). The light green shaded region shows the uncertainty on this prediction because we have measured $\alpha$ from the simulation which has a finite number of pairs.
 The AAP function plotted here was found assuming a $\Lambda$CDM cosmology for $H(z)$.
The error bars on the data points are estimated from a jack knife sampling of the subhalo pairs using 100 subsamples of the data.
}
\label{5a}
\end{figure}

In Fig. \ref{5a} we plot the measured mean of each of the distributions shown in Fig. \ref{pdf} as a function of redshift. The red dot-dashed line shows the mean 
of the distribution of all subhalo pairs in redshift space with no cuts. By restricting the sample, 
using either a cut in  $\Delta v$ or $\theta$, we obtain the  mean shown as the green dot-dot-dashed line 
or orange dotted line respectively.
Selecting subhalo pairs according to the two selection cuts discussed above results in a measured mean (blue dashed line) which is in very 
good agreement with the predictions of the corresponding AAP function (solid black line) when we
measure $\alpha$ directly from the simulation.
Note each distribution has its own associated AAP function, with a normalization set by the pairs in each sample.

The error bars in Fig. \ref{5a} have been calculated by jackknife sampling the subhalo pairs by grouping the data into 100 
sets containing equal numbers of subhaloes, and then successively removing one set at a time, calculating the sample mean for the remaining haloes and
computing the variance amongst the measured means \citep[see][for a discussion of the reliability of the jackknife technique]{2009MNRAS.396...19N}. 
We have verified that these errors change the AAP function by less  1\%   if  we vary the sample size to 25 or 50 subsamples at a given redshift.
The error on the AAP function, shown as a green shaded region in Fig. \ref{5a}, was found using a similar method to find the variance in $\alpha$ at each
redshift splitting the pairs in the simulations into 100 subsamples.
 The errors on both $\langle \sin^2 \tau \rangle$ and on the AAP function increase with increasing redshift as the number of pairs decreases.
This happens because of the fixed resolution of the simulation, which means that we resolve a progressively smaller
fraction of the subhalo population with increasing redshift. A similar drop in the number of pairs would happen in a flux limited galaxy survey.

\begin{figure}
{\epsfxsize=8.5truecm
\epsfbox[56 366 366 650]{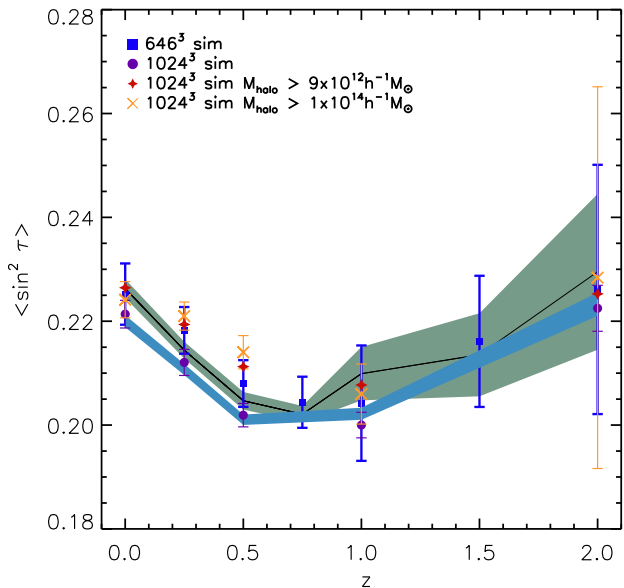}}
\caption{ The sample means of the anisotropic distribution of pairs of subhaloes measured
in the low and high resolution $\Lambda$CDM simulations are shown as blue squares and purple circles respectively. The AAP function given in Eq. \ref{aapfunction} using the value for
$\alpha$ measured from the simulation at each redshift is shown as a light green and a
blue shaded region for the low and high resolution simulation respectively. 
The mean measured using subhalos from the higher resolution simulation that have a parent halo mass of $M \ge 9 \times 10^{12} h^{-1}M_{\sun}$ 
and $M \ge 1 \times 10^{14} h^{-1}M_{\sun}$ are shown as red stars and orange crosses respectively. Note that for the measured mean of subhalos within a parent halo of $M \ge 1 \times 10^{14} h^{-1}M_{\sun}$ at $z=2$ there are only 49 pairs and we used 10 subsamples to find the jackknife errors.
}
\label{3b}
\end{figure}

We have  tested the stability of the method by comparing  simulations of different resolution. 
The  sample mean,  $\langle \sin^2\tau \rangle$ (Eq. \ref{sin2t}), of the distribution of subhalo pairs in redshift space  in the
two $\Lambda$CDM simulations, higher and lower resolution,
is shown in Fig. \ref{3b} at different redshifts.
The mean for the lower  resolution simulation is shown as blue squares in Fig. \ref{3b}. 
The AAP function using the measured value for $\alpha$ at each redshift is shown as a solid black line as in Fig. \ref{5a}.
The sample mean from the higher resolution simulation is shown as purple circles with the corresponding AAP function shown as a solid light blue line.
In both the lower and higher resolution simulations, the measured distribution of pairs shows excellent agreement with the predictions of the AAP function assuming a $\Lambda$CDM cosmology,
 and also agree with each other in shape and amplitude
within the error bars.
The difference between the AAP function for the higher resolution (blue shading) and for the lower resolution simulation (solid black line) in Fig. \ref{3b} is also due to the difference in resolution between the two simulations.
If we select only subhalos from the higher resolution simulation that have a halo mass of $M \ge 9 \times 10^{12} h^{-1}M_{\sun}$, which corresponds to the minimum halo mass
  selected by the FOF algorithm in the lower
resolution simulation, we obtain the red stars with errors bars plotted in Fig. \ref{3b}.  
These points are almost coincident with the corresponding measurement from the lower resolution simulation (blue squares), agreeing to better than 1$\sigma$.

We also make contact with an observational galaxy sample in Fig. \ref{3b}.
If we select subhalos from main halos which have a mass of $M \ge 1 \times 10^{14} h^{-1}M_{\sun}$ we obtain the mean plotted as orange crosses in Fig. 
\ref{3b}. 
Again these results are consistent with the means for the lower resolution  simulation at each redshift.
This mass corresponds to the minimum halo mass expected to contain two or more luminous red galaxies  (LRGs) 
on average \citep{2008MNRAS.386.2145A}. This subhalo selection is relevant for a
spectroscopic redshift survey such as the SDSS-III BOSS survey \citep{2007AAS...21113229S} which will target LRGs in the redshift range $z<0.7$.
Without applying any selection cuts, we find approximately 27,000 subhalo pairs at $z=0$  which share a common  halo of $M \ge 1 \times 10^{14} h^{-1}M_{\sun}$, at $z=0.25$ and $z=0.5$ the number of subhalo pairs are approximately
18,000 and 11,000 respectively. From the first semester of BOSS data, \citet{2011ApJ...728..126W} estimate 
that the cumulative probability that a galaxy in their sample is hosted by a halo 
of mass $M \ge 1 \times 10^{14} h^{-1}M_{\sun}$ is about 5\%. If we extend this probability to the full sample of
LRGs expected by BOSS with  space density $\bar{n} = 3 \times 10^{-4}h^3/$Mpc$^3$, this corresponds to approximately 13,000 pairs of LRGs in the redshift range $z=0.5-0.6$.  This is  
similar to the number  of pairs
we obtain from the higher resolution simulation restricting to halos with to $M \ge 1 \times 10^{14} h^{-1}M_{\sun}$ at $z=0.5$, shown by the orange crosses in Fig. \ref{3b}.

 The errors on the AAP function as measured by  Marinoni \& Buzzi, $\alpha = 5.79^{+0.32}_{-0.35}$, are substantially larger than ours due to the uncertainty in fitting for the parameter
$\alpha$ at $z=0$ with a smaller number of pairs. 
Our higher resolution simulation has approximately 4 times more subhalo pairs
than the lower resolution simulation, after making the selection cuts discussed in Section 4, which gives rise to error bars which are
approximately 50\% smaller in the higher resolution run (see Fig. \ref{3b}). The sample of pairs used by  Marinoni \& Buzzi is approximately 25 times smaller
then the sample from our lower resolution simulation. 

We have verified that by applying the Marinoni \& Buzzi selection cuts to our parent sample of subhalo pairs in the lower resolution simulation gives 
$\alpha = 5.69$, which is consistent with the value for $\alpha$ obtained by these authors. However, we find that 
the measured mean for this simulation sample does not agree 
with the AAP function within the error bars (if our sample was the same size as that used by Marinoni \& Buzzi our errors would be significantly larger 
and the two would agree in this case). 
This demonstrates the need for the robust resolution independent selection criteria we have presented here.

\section{Application: a new test \label{section3}}

In this section we use the selection criteria outlined in Section 4 to test the predictions of the AAP function, Eq. \ref{aapfunction}, using the distribution of subhalo pair angles measured in N-body simulations.

The accuracy of this test relies on two key variables: the cosmological expansion history assumed, $H(z)$ and the normalization parameter, 
$\alpha = H^{-1}_0(\langle \Delta v^2_{\parallel}\rangle/\langle \Delta r^2\rangle)^{1/2}$.
We consider the impact of uncertainties in each of these variables in turn.
In Section \ref{sub3.1} we present the measured anisotropic
distribution of the orientation of pairs, selected according to the prescription set out in Section 4, and its first moment at different redshifts
 together with the predicted distribution using the AAP function in
a $\Lambda$CDM and in two quintessence dark energy cosmologies.
In order to test the ability of the theoretical model to distinguish different cosmologies we will 
assume perfect knowledge of the correct $H(z)$ and $\alpha$ in the first instance. We then consider how an observer would measure $\alpha$ and the impact this has on the results, again assuming the
correct $H(z)$.
We relax the  assumptions further in
 Section \ref{sub3.2}  where an incorrect cosmological expansion history is used to analyse the data. This is done by
 measuring the distribution of subhaloes in the INV  and SUGRA dark energy simulations  assuming a
$\Lambda$CDM cosmology to infer distances to the pair.
We will show that the method, as implemented in \citet{2010Natur.468..539M}, fails to exclude the wrong cosmology. 
Consequently, we propose a new method, which uses the theoretical model  discussed so far but which exploits additional 
information about $\alpha$ from the numerical simulations. In Section 5.3 we show that this method can be successfully applied to test dark energy.

\subsection{Testing the method: theory versus observations \label{sub3.1}}

\begin{figure}
{\epsfxsize=8.5truecm
\epsfbox[56 366 366 650]{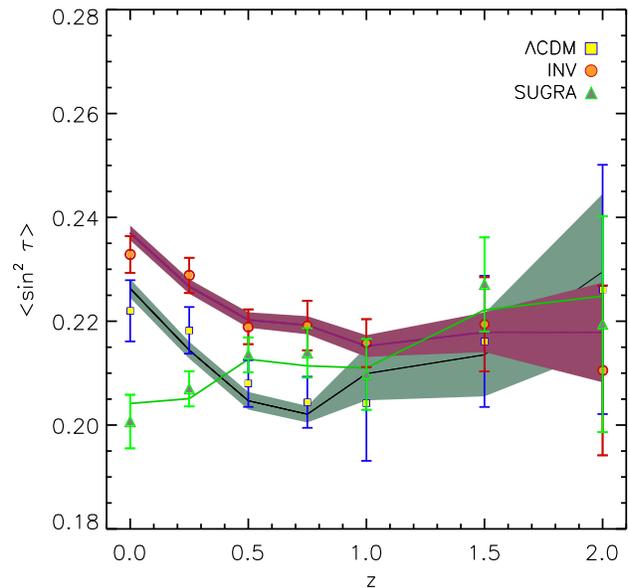}}
\caption{ The first moment of the anisotropic distribution of pairs of subhaloes measured in a $\Lambda$CDM  and two quintessence dark energy simulations as a function of redshift.
Measurements for $\Lambda$CDM,  INV and SUGRA cosmologies  are shown as blue squares, red circles and green triangles  respectively. The AAP function using
 the  measured value for $\alpha$ at each redshift for each cosmological model is shown as solid black, red and green lines for $\Lambda$CDM, SUGRA and INV 
respectively. The shaded bands show the uncertainty on the AAP function for each cosmology.
Note that the error bars for the AAP function for SUGRA are similar to those for the INV model and are not shown for clarity.
}
\label{lcdminvsugra}
\end{figure}
First of all, we test how the approach discussed in Section 2 can distinguish different cosmologies. We put ourselves in the idealised situation of an observer who knows the correct 
cosmological model to compute distances and is able to measure peculiar velocities precisely to find $\alpha$ at each redshift.
In Fig. \ref{lcdminvsugra} the mean of the redshift space distributions of subhalo pairs for $\Lambda$CDM and the two quintessence dark energy models are plotted as 
a function of redshift. The results for $\Lambda$CDM are the same as those shown in Fig. \ref{3b} for the lower resolution simulation. 
The measured sample mean for the INV dark energy model is shown as red-orange circles with error bars while the results for the SUGRA model are shown as green-grey triangles.
The predicted AAP function for each of these models, using the correct expansion history and the value for $\alpha$ measured at each redshift, is shown as 
a solid red line for the INV model and a solid green line for the SUGRA model.  The uncertainties on the AAP function are
 plotted as a red shaded region for the INV model. The errors for the SUGRA model are similar but are not plotted in Fig. \ref{lcdminvsugra} for clarity.
The errors shown on both models for the measured mean and the AAP function were found using an identical jackknife sampling method to that used for the $\Lambda$CDM result.
It is clear from Fig. \ref{lcdminvsugra} that the  measured mean for the three simulations agree with the respective AAP function, provided the correct 
expansion history is known and that the parameter $\alpha$ can be determined at each $z$. 
As these results show,
 the measurements in a $\Lambda$CDM or a dynamical dark energy model  agree very well with the predictions, if the correct cosmology is used to analyse the data.
For the two quintessence models the deviations from $\Lambda$CDM are due to the different expansion histories (see Fig. \ref{H}).
This is a consistency check which confirms that the method works.

In reality, in a galaxy survey, it is not possible to measure the parameter $\alpha$ accurately at high redshifts because of the difficulties associated with measuring galaxy peculiar velocities 
to sufficient precision. 
We shall now degrade the status of the idealised observer mentioned above and consider a more realistic observer who still knows the correct cosmological model 
but who is unable to measure $\alpha$ directly at any redshift other than $z=0$.
Using the measured distribution of pairs at $z=0$, we fit 
the distribution given in Eq. \ref{dist} to set $\alpha$ and test the accuracy of the AAP function using this $\alpha(z=0)$ value at each redshift, as suggested by   Marinoni \& Buzzi. 
If $\alpha$ does not evolve with redshift we would expect this approach to result in accurate agreement between the measured mean and the AAP function.

\begin{figure}
{\epsfxsize=8.5truecm
\epsfbox[56 360 323 614]{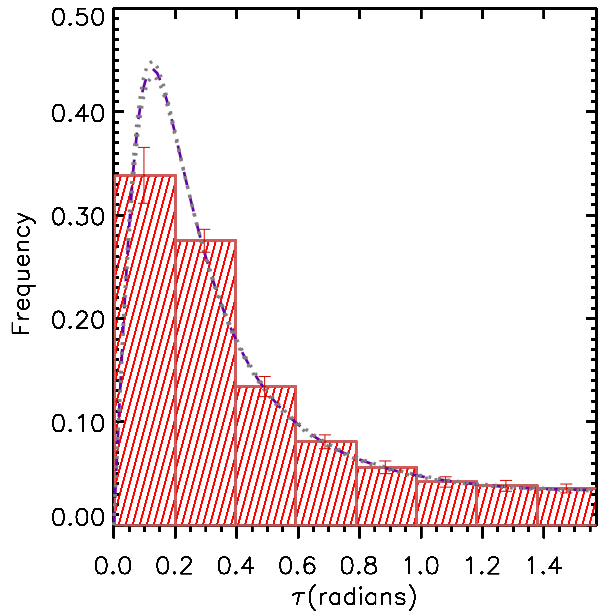}}
\caption{ The fractional distribution of the angle $\tau$ in radians of pairs of subhaloes measured in $\Lambda$CDM  at $z=0$.
The error bars on each bin are calculated by jackknife sampling after dividing the catalogue of subhalo pairs into 100 subsamples and calculating the variance amongst the distributions measured after successively
removing one subsample at a time.
The purple dashed line shows the distribution in Eq. \ref{dist}, with the best fit value of the normalization parameter $\alpha_{\mbox{\tiny FIT}} = 5.67 \pm 0.1$. The grey dotted lines show the 1 $\sigma$ error on the best fit distribution.
} \label{fit_alpha}
\end{figure}

\begin{figure}
{\epsfxsize=8.5truecm
\epsfbox[56 366 366 650]{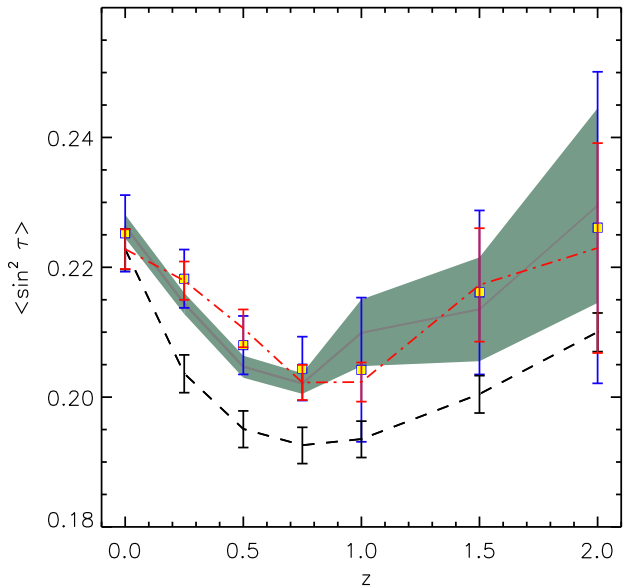}}
\caption{ The first moment of the anisotropic distribution of pairs of subhaloes measured in a  $\Lambda$CDM simulation (blue-yellow squares) as shown in Fig. \ref{lcdminvsugra}.
The AAP function using the $z=0$ best fit value for $\alpha$ at each redshift (see Fig. \ref{fit_alpha}) is shown as a
dashed black line with error bars. If we fit for $\alpha$ using the measured distribution at each redshift we obtain the red dashed line prediction for the AAP function assuming a $\Lambda$CDM cosmology.
The error bars on the black dashed and red dot-dashed line are the 1 $\sigma$ errors obtained by fitting for $\alpha$.
}
\label{plot_alpha}
\end{figure}

In Fig. \ref{fit_alpha} the measured distribution of the angle $\tau$, in radians, for 
$\Lambda$CDM  is shown as a red hashed region with error bars. Note the y-axis shows the fraction of the 
total number of pairs per bin. The distribution (Eq. \ref{dist}) with best fitting value for $\alpha_{\mbox{\tiny FIT}} = 5.67 \pm 0.1$ (with 
1$\sigma$ errors) is plotted 
as a purple dashed line. The grey dotted lines show the distribution adopting $\alpha + 1\sigma$ and $\alpha - 1\sigma$. Note the error we obtain for $\alpha$, 0.1,  is much smaller than
that obtained by Marinoni \& Buzzi (0.3) due to the difference in sample size and the different methods used to estimate the errors.
 This value for  $\alpha$ agrees with the measured value from the simulations of $\alpha = 5.56$.
In Fig. \ref{plot_alpha} the AAP function assuming a $\Lambda$CDM cosmology and using this $\alpha_{\mbox{\tiny FIT}}(z=0)$ 
value  at each redshift
is shown as a black dashed line with error bars. 
The mismatch between this curve and the simulation results clearly indicates that $\alpha$ evolves with redshift,  
invalidating one of the main assumptions made in the analysis of
Marinoni \& Buzzi.
Note that this black dashed line in Fig. \ref{plot_alpha} is much smoother than the shaded green band 
for the AAP function in $\Lambda$CDM where the value of $\alpha$ is measured 
directly from the simulation at each redshift.
Using the $z=0$ value for $\alpha$ produces an AAP function which is systematically and significantly below the measured results for a $\Lambda$CDM 
cosmology for $z>0$. Applying this measure of $\alpha$ as proposed by Marinoni \& Buzzi could lead to a spurious detection of deviations from $\Lambda$CDM. 

It is clear from  Fig. \ref{plot_alpha} that the method proposed by Marinoni \& Buzzi contains a serious systematic error which is apparent when applied to 
a large sample of pairs. Marinoni \& Buzzi considered a smaller sample than in the simulations where the statistical errors dominated this systematic.

It is clear from Fig. \ref{plot_alpha} that $\alpha$ does evolve with redshift and that we can improve on the estimates of this parameter by fitting Eq. \ref{dist} to the measured distribution at each redshift.
In Fig. \ref{plot_alpha}  the AAP function in a $\Lambda$CDM cosmology using the best fit values for $\alpha$ measured at each redshift is shown as a red dot-dashed line with error bars.
The jackknife errors on $\alpha$ are estimated using 100 subsamples for the distributions at $z=0 - 1$ and 50 subsamples for $z=1.5$ and 2 as there are fewer pairs at these higher redshifts.
This approach to measuring $\alpha$ gives much better agreement with the mean measured from the simulations, shown as blue-yellow data points in Fig. \ref{plot_alpha}.
Note this method of extracting $\alpha$ assumes that the correct cosmology is $\Lambda$CDM.

\subsection{The test assuming a particular cosmology \label{sub3.2}}

In this section we are no longer idealised observers who know the correct cosmology, 
so the only possible choice is to assume the same cosmology in the data fitting and in the theoretical prediction of the galaxy distribution.
Specifically we will  assume
$\Lambda$CDM, for simplicity, 
in order to set the expansion history $H(z)$ in Eq. 8 and to compute the comoving distances in Eq. \ref{sin2t}, as well as to extract the parameter $\alpha$.
In order to find the parameter $\alpha$ we must fit to the observed distribution of the orientations of pairs which has been found also by assuming a $\Lambda$CDM cosmology.
Assuming that the true cosmological model chosen by nature is a dynamical dark energy model, for 
instance the INV or SUGRA cosmology, we will check if the wrong cosmology, $\Lambda$CDM in our case, can be excluded or not, and 
consequently if the method is applicable to future galaxy surveys.
For this analysis we take subhalo pairs in the INV and SUGRA simulations and at each redshift we rescale the comoving distances to match those which would be computed by an 
observer assuming a $\Lambda$CDM cosmology. 

\begin{figure}
{\epsfxsize=8.5truecm
\epsfbox[56 366 366 650]{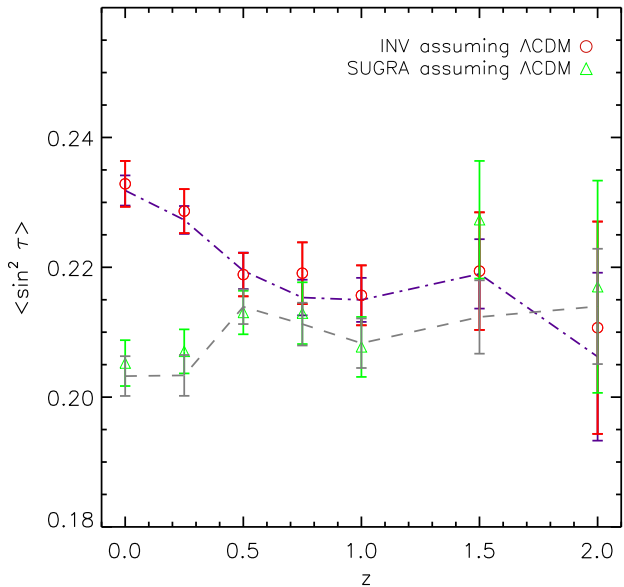}}
\caption{ The measured  mean of the anisotropic distribution of pairs of subhaloes in the  INV dark energy simulation {\emph{assuming}} a $\Lambda$CDM cosmology to find the comoving distance to each pair member (red circles).
The purple dot-dashed line shows the predicted AAP function, assuming a $\Lambda$CDM cosmology, and using the best fit $\alpha$ found at each redshift.
The measured mean for the SUGRA model, {\emph{assuming}} a $\Lambda$CDM cosmology, is shown as  green triangles and the predicted AAP function, assuming a $\Lambda$CDM cosmology, and using the best fit $\alpha$ found 
at each redshift is shown as a grey dashed line.
}
\label{assumelcdm}
\end{figure}

The ensemble average of Eq. \ref{sin2t} for each subhalo pair in the INV (SUGRA) simulation is plotted in 
Fig. \ref{assumelcdm} as  red circles (green triangles), with error bars. 
If we fit for $\alpha$ at each redshift, we obtain the purple dot-dashed line in Fig. \ref{assumelcdm} for the INV model. 
Although we have assumed,  incorrectly a $\Lambda$CDM cosmology we find that the AAP function agrees with 
the measured sample mean for the INV model at each redshift. 
In Fig. \ref{assumelcdm} a similar analysis is presented for the SUGRA model.
The measured mean for this dark energy model, assuming a $\Lambda$CDM cosmology to compute comoving distances, is shown as  
green triangles. The AAP function using the best fit value for $\alpha$ at each redshift and a $\Lambda$CDM expansion history is shown as a grey dashed line.
Again, theory and observations agree when we would expect them not to as we have used the  wrong cosmology in the AAP function and to compute distances.
Our results show that the AAP function, using either a fixed value of $ \alpha(z=0)$ or a best fit value at each redshift,
is not an accurate model with which to test for dynamical dark energy models if the correct cosmological model is unknown and that further input from numerical simulations is needed to arrive at a viable test.

\subsection{A blueprint for probing dark energy}

From the previous section it is clear that the cosmological test 
proposed by Marinoni \& Buzzi relies heavily on measuring the parameter $\alpha$ accurately at each redshift.
The assumption that $\alpha$ does not vary with redshift is incorrect and could falsely rule out $\Lambda$CDM if this test is misapplied to pairs of galaxies in future surveys.
The value of $\alpha$ also depends on the cosmological model. For example at $z=0$ the values for the $\Lambda$CDM, SUGRA and INV cosmologies are $\alpha=5.56$, $\alpha=6.32$ and 
$\alpha=5.32$ respectively, with a typical error of 0.1. The difficulty is not just a problem of measuring $\alpha$ accurately but stems from the fact that the cosmology assumed affects both the data and the 
theoretical prediction in a way which cannot be disentangled. 
The accuracy and predictive power of the AAP function can be restored if instead of measuring $\alpha$ from 
observations, we employ N-body simulations which contain a comparable number of subhalo pairs to the number of 
galaxy pairs in the survey under consideration.
It is clear that independent information about $\alpha$ is necessary and numerical simulations play an important role in providing 
these predictions in a given cosmology.
We propose a new approach to measuring dark energy, where observational measurements of the mean of the anisotropic distribution of pairs and predictions of the AAP function from numerical simulations are combined.

The new method we propose to test a given cosmology is as follows:

\begin{itemize}
\item An observer assumes the cosmology to obtain the comoving distances needed to calculate the ensemble average of Eq. \ref{sin2t} for a sample of pairs of galaxies, selected using the criterion in 
Section 4, at different redshifts. 
\item Using a  N-body simulation of the same assumed cosmology and with a comparable number density of pairs and volume to the galaxy survey, the observer can then construct a similar catalogue of pairs according to Section 4 and find the value of $\alpha$ at each redshift. 
\item This gives rise to a prediction for the AAP function which can be compared with the means measured from 
the galaxy survey at each redshift and the assumed cosmology can be verified or  excluded.
\end{itemize}

Note if the AAP function measured from the simulation and the measured mean of the galaxy pair sample, analysed assuming the same cosmology disagree, then 
a suite of N-body simulations of different cosmologies would need to be run. The AAP function from each
 simulation should then be compared to the measured mean, computed  assuming the cosmology used in the simulation. This test 
is realistic given current computing resources.

\begin{figure}
{\epsfxsize=8.truecm
\epsfbox[58 362 259 650]{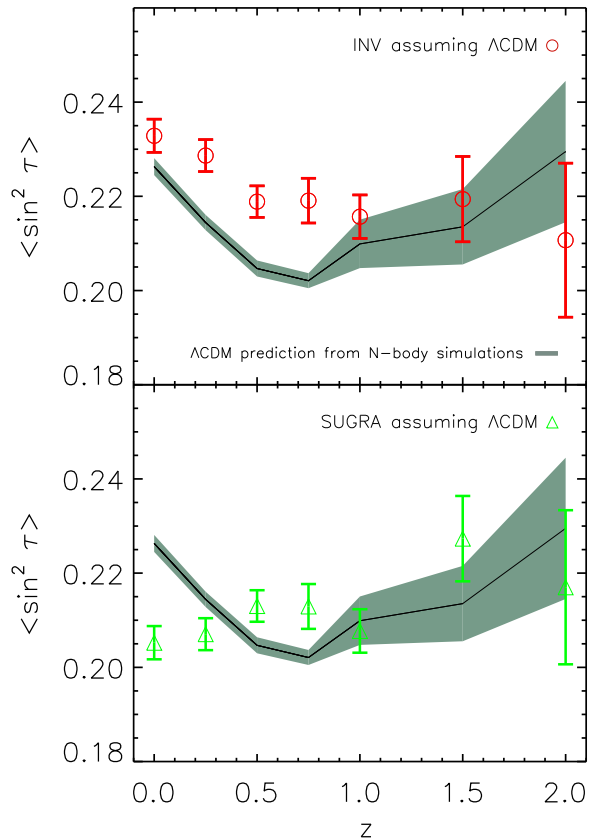}}
\caption{ Upper panel: The measured  mean of the anisotropic distribution of pairs of subhaloes in the  
INV dark energy simulation {\emph{assuming}} a $\Lambda$CDM cosmology to find the comoving distance to
 each pair member (red circles). The predicted AAP function  for a $\Lambda$CDM cosmology, using the value of $\alpha$ measured directly from the lower resolution N-body simulation, is shown as a solid black line.
Lower panel: same as the upper panel but for the SUGRA quintessence model assuming a $\Lambda$CDM cosmology to determine comoving distances (green triangles).
}
\label{assume2}
\end{figure}

In Fig. \ref{assume2}  we use the INV and SUGRA simulations to illustrate this  method.
In the upper and lower panels we show the measured means for the INV and the SUGRA  dark energy simulations respectively, 
which are treated here as the \lq observed\rq \, pair sample. 
In this example we are testing a $\Lambda$CDM cosmology and use it to compute the distances in 
each case, as in Fig. \ref{assumelcdm}, together  with the predicted AAP function from an N-body 
simulation of $\Lambda$CDM  where $\alpha$ is measured directly from the simulation (green shaded region). 
It is clear that for $z<1$ the INV model and the SUGRA model can be distinguished from the AAP function predicted in a $\Lambda$CDM
cosmology. This result shows that if a SUGRA or INV model is the correct cosmology for the Universe then $\Lambda$CDM can be ruled out. 
If there is a mismatch between the measurement from the observed pair sample and the simulation calibrated AAP prediction as in Fig. \ref{assume2}, then a new simulation with a different expansion history is required until
an acceptable match is found.

\section{Summary}

The distribution of the orientation of pairs of galaxies is uniform in real space in a homogeneous and isotropic universe. 
However in redshift space, two effects lead to the inferred distribution
becoming skewed. Firstly, an observer has to assume a cosmology to
convert positions on the sky and redshifts into distances. A mismatch
between the assumed and underlying cosmology introduces an error in the
radial distance to a galaxy. Secondly, peculiar motions introduce distortions 
which break the connection between the  measured redshift
and the actual distance. Both effects result in an apparent displacement of
galaxies along the line of sight.

\citet{2010Natur.468..539M} proposed that the distortion of the
distribution of the angle subtended between galaxy pairs as
viewed in redshift space can be modelled by a simple
Doppler shift in the galaxy positions. 
 This procedure gives rise to a theoretical prediction for
the distribution in
redshift space, which is referred to as the AAP function. We have
tested the accuracy of this model using subhalo pairs identified in N-body
simulations of cosmologies with different dark energy models.

The AAP function depends on two variables:
the ``normalization'' parameter $\alpha =
H^{-1}_0(\langle \Delta v^2_{\parallel}/\Delta r^2\rangle)^{1/2}$, and
the expansion history, $H(z)$, which depends on the cosmology.
In this paper we present the AAP function
normalized in three different ways: (i) using the relation $\alpha =
H^{-1}_0(\langle \Delta v^2_{\parallel}/\Delta r^2\rangle)^{1/2}$,
we can measure $\alpha$ directly from the simulation at each redshift,
(ii) we can measure $\alpha$ at $z=0$ by fitting to the measured
distribution and then, assume that $\alpha$ does not evolve with redshift,
and apply the  $z=0$ normalization to specify the mean of the distribution
of pairs at different redshifts (as suggested by  Marinoni \& Buzzi),
(iii) we can apply case (ii) but fitting  for $\alpha$ at each redshift using the measured distribution and not just at $z=0$.
When we measure $\alpha$ directly (case i), we obtain excellent
agreement between measurements of the mean from the
simulation and the predictions of the AAP function.
If instead we retain the best fit $z=0$ value, $\alpha(z=0)$,
at each redshift (case ii), we do not find a good match between
the theory and the simulation measurements.
 This demonstrates that simply fitting for $\alpha$ at $z=0$ and
assuming it does not evolve with redshift is not accurate.
 In fact such an
approach would incorrectly rule out the cosmology used in the simulation.
If we fit for $\alpha$ at each redshift using the simulation (case iii),  then we again recover
an excellent match between the theory and simulation results. 

We use a large sample of subhalo pairs which do not necessarily reside in 
fully relaxed and virialised haloes that have detached from the Hubble flow.
This is demonstrated by the fact that we measure a different value for $\alpha$ in different cosmologies, see Section 5.

Note that each of
the above cases consider idealised
observers who know the correct cosmology to compute distances and $H(z)$.
The measured mean of the distribution of pair angles (Eq. \ref{sin2t})
depends on the cosmological model assumed to convert position on the
sky and redshift to comoving distance.
The AAP function also depends on cosmology through $H(z)$.
As a result, the measured mean and the AAP function will not agree
if the wrong cosmology is assumed (the Alcock-Paczynski  effect).
Using two quintessence dark energy simulations (labelled INV and SUGRA),
we have tested if the AAP function reproduces the measured mean
of the distribution when, in the first instance, we know the correct cosmology
(the \lq perfect\rq \, observer case), and in the second instance,
when we instead assume $\Lambda$CDM (i.e. the \lq real\rq \, observer who
has no prior knowledge of the underlying cosmology). The two dark
energy models we consider have an evolving equation of state
which is compatible with current observations of the CMB, BAO and
Type Ia SN distances. We find that, for a perfect observer who knows
$H(z)$ and $\alpha$ exactly, the AAP function and the measured means
are in very good agreement for both the SUGRA and the INV models.

Consider now performing the same exercise using the SUGRA and INV
simulations, as a real observer who does not know the underlying
cosmology and so assumes a $\Lambda$CDM cosmology, and who uses the best
fit value for $\alpha$ at each redshift. We might expect that the
theory should not match the measured mean for the  dynamical dark energy
models. However, we find that, by fitting for $\alpha$ using the observed 
distribution in the simulations, we instead recover a model which incorrectly 
matches the observations extremely well for both dark energy cosmologies, 
even though we have assumed a $\Lambda$CDM model.  The consequences are 
that, in a universe with evolving dark energy, we would find that a $\Lambda$CDM 
model incorrectly matches the observations invalidating the methodology.

In this paper we have proposed a new formulation of the test of Marinoni \& Buzzi 
in which the distribution of galaxy pairs can be analysed without prior knowledge 
of the cosmology.
The measured distribution of angles should  be compared with 
predictions for the AAP function using a reference N-body simulation 
to directly measure $\alpha$.  We have shown that the subhalo pairs in 
two quintessence dark energy simulations, which are treated as the ``observed'' pair sample 
in this instance, produce a different measured  distribution to that predicted in a
$\Lambda$CDM simulation even when analysed after assuming (incorrectly) 
a $\Lambda$CDM cosmology. In the new test, the AAP function is normalized with 
reference to a simulation with the same cosmology assumed to analyse the observations. 
The predicted AAP function and measurement will only agree if the assumed cosmology matches 
the true cosmology. If this is not the case, then a new reference simulation must be 
generated with a revised expansion history, to see if an improved match to the observed 
distribution of galaxy pair angles can be obtained. 
We find that, by measuring the mean of the distribution as a function of redshift, we should 
be able to detect deviations from a $\Lambda$CDM  expansion history at the level of 2\% in 
a box of volume $\sim 3 h^{-3}$Gpc$^3$.
This new test complements the constraints on the present value of Hubble's
parameter provided by observations of Type Ia SN which constrain $H_0$
to $\sim 3$\% \citep{2011ApJ...730..119R} and improves on constraints
of $H(z)$ at higher redshifts which are accurate to about $\sim 10$\%.


\section{Conclusions}

Distinguishing between competing scenarios for the accelerating
expansion of the universe is a major challenge for both
observational and theoretical cosmologists. The expansion history and
distance-redshift relations are remarkably close between viable
models which satisfy the currently available constraints.
A convincing determination of the nature of dark energy will
require a combination of probes for two reasons \citep{2006astro.ph..9591A}. Firstly, the small
differences in the expected signals from a given probe mean that
systematic effects become important. Applying different probes
will allow us to see whether or not a measured signal is robust
to systematics. Secondly, some existing tests cannot distinguish between
some classes of dark energy model. New probes are therefore needed
to break such degeneracies.

We have tested one such example of a new probe, the
distribution of angles subtended between pairs of galaxies.
This distribution is distorted by the peculiar motions of
galaxies and  also by the choice of cosmology adopted to
transform observed positions into comoving distances.
The origins of this test can arguably be traced back to
Alcock \& Paczynski (1979), and it was refined by  \citet{ 1994MNRAS.269.1077P}. 
\citet{2010Natur.468..539M} applied the test to the angle between 
pairs of galaxies and crucially included redshift space distortions.

We have used numerical simulations of structure formation to
assess the performance of the test. The mean of the distribution
of pair angles varies with redshift and, furthermore, is measurably
different between cosmologies. A comparison  between a theoretical
model for the pair angle distribution and the measurements from
the simulations shows that the test, as originally proposed, is
limited. The theoretical calculation requires a parameter to be
specified to normalize the distribution of pair angles. Our simulations
show that this parameter is redshift and cosmology dependent.

It is possible to estimate the normalization of the pair angle
distribution observationally, at redshifts $z>0$, if the peculiar
velocities of galaxies can be measured.  For example, it was
recently argued that accurate mean pairwise velocities  of
pairs of Type Ia SN can be obtained by combining photometry from
a survey such as Pan-STARRS \citep{2010SPIE.7733E..12K}
or the Large Synoptic Survey Telescope \citep{2009arXiv0912.0201L}
with follow-up spectroscopy \citep{2011PhRvD..83d3004B}.
At present the accuracy of measurements of the peculiar velocity
field is not adequate to distinguish between the models
compared in this paper.

Our proposed methodology avoids this problem by using an N-body
simulation with a similar number of pairs to the observational
sample to normalize the distribution of angles. This secures
the crucial step of setting the normalization of the theoretical
distribution at each redshift. The detailed selection of the
N-body sample of subhalo pairs is not important, avoiding the need
to combine the simulation with a galaxy formation model.
Furthermore, we have demonstrated that it is not necessary to
have a knowledge of the true underlying background cosmology for
the successful application of the test.

The new method we have proposed is a powerful complement and extension
to existing probes of dark energy. This is demonstrated by the
ability of the pair distribution to distinguish between
cosmologies that cannot be separated through the appearance of
BAO or through the halo mass function. The technique can be
applied already to ongoing surveys, such the SDSS-BOSS survey
\citep{2007AAS...21113229S}, and should yield competitive constraints.
The method should also produce distinct signals for dark energy
and modified gravity models which have identical expansion histories,
through the different peculiar motions induced.

\section*{Acknowledgments}

EJ acknowledges receipt of a fellowship funded by the European
 Commission's Framework Programme 6, through the Marie Curie Early Stage
Training project MEST-CT-2005-021074.
This work was supported in
part by grants from the Science and Technology Facilities
Council held by the Extragalactic Cosmology Research Group and 
the Institute for Particle Physics Phenomenology at Durham University.
SP thanks the Fermilab Theoretical Physics Department for hospitality.
The calculations for this paper were performed on the ICC 
Cosmology Machine, which is part of the DiRAC Facility jointly funded by STFC, 
the Large Facilities Capital Fund of BIS, and Durham University.
We thank the referee and Will Percival for helpful comments.

\bibliographystyle{mn2e}
\bibliography{mybibliography}

\bsp

\label{lastpage}

\end{document}